\documentclass[preprint,showpacs,preprintnumbers,amsmath,amssymb]{revtex4}
\usepackage{amsfonts}
\usepackage{amssymb}
\usepackage{latexsym}
\usepackage{graphicx}

\newcommand{\al}{\alpha}
\newcommand{\tal}{\tilde \alpha}
\newcommand{\be}{\beta}
\newcommand{\tbe}{\tilde\beta}
\newcommand{\pa}{\partial}

\newcommand{\si}{\sigma}

\newcommand{\de}{\delta}

\newcommand{\tha}{\theta}

\newcommand{\rar}{\rightarrow}

\newcommand{\non}{\nonumber}
\begin{document}

\preprint{Preprint M\'exico ICN-UNAM 03-10}

\title{The H$_2^+$ ion in a strong magnetic field. \\
 Lowest excited states }

\author{A.~V.~Turbiner}
 \altaffiliation[]{On leave of absence from the Institute for Theoretical
 and Experimental Physics, Moscow 117259, Russia}
\email{turbiner@nuclecu.unam.mx}
\affiliation{%
Instituto de Ciencias Nucleares, Universidad Nacional
Aut\'onoma de M\'exico, Apartado Postal 70-543, 04510 M\'exico, D.F., Mexico.}
\author{J.~C.~L\'opez Vieyra}
\email{vieyra@nuclecu.unam.mx}
\affiliation{%
Instituto de Ciencias Nucleares, Universidad Nacional
Aut\'onoma de M\'exico, Apartado Postal 70-543, 04510 M\'exico, D.F.,
Mexico.}

\date{October 29, 2003}

\begin{abstract}
  As a continuation of our previous work ({\it Phys. Rev. A68, 012504 (2003)})
  an accurate study of the lowest $1\si_g$ and the low-lying excited
  $1\si_u$, $2\si_g$, $1\pi_{u,g}$, $1\de_{g,u}$ electronic states of the
  molecular ion $H_2^+$ is made. Since the parallel configuration where
  the molecular axis coincides with the magnetic field direction
  is optimal, this is the only configuration which is considered.
  The variational method is applied and the {\it same} trial
  function is used for different magnetic fields.
  The magnetic field ranges from $10^9\,G$ to $4.414 \times
  10^{13}\,G$ where non-relativistic considerations are justified.
  Particular attention is paid to the $1\si_u$ state which was studied
  for an arbitrary inclination. For this state a one-parameter vector potential
  is used which is then variationally optimized.

\end{abstract}

\pacs{31.15.Pf,31.10.+z,32.60.+i,97.10.Ld}

\maketitle

\section{Introduction}

In our previous paper \cite{TV:2003} (cited below as I) we carried
out an accurate detailed study of the ground state $1_g$ of the
molecular ion $H_2^+$ placed in a constant uniform magnetic field
ranging from zero up to $4.414 \times 10^{13}\,G$ for all
inclinations $0^o - 90^o$. The goal of that study was to
investigate the domain of existence of the $H_2^+$ ion. We showed
that for all magnetic fields studied the molecular ion $H_2^+$
exists for moderate (not very large) deviations of the molecular
axis from the magnetic field direction (moderate inclinations).
Furthermore it was found that for each magnetic field the most
stable configuration of minimum total energy corresponded to zero
inclination, where the molecular axis coincides with magnetic
field direction. We called this configuration the `parallel
configuration'. To this configuration the standard spectroscopic
notation $1\si_g$ can be assigned. A major feature of this
configuration is that with magnetic field growth the system
becomes more and more bound (binding energy grows) and more and
more compact (equilibrium distance decreases).

The aim of the present paper is to continue the study initiated in
I and to explore several low-lying excited states. At first we
re-examine the ground state for the parallel configuration
$1\si_g$ in the region $10^9 - 4.414 \times 10^{13}\,G$. A
detailed study of the $1\si_u$ state which is anti-bonding without
a magnetic field is presented. Then the lowest states of different
magnetic quantum numbers are investigated as well as the $2\si_g$
state.

Atomic units are used throughout ($\hbar$=$m_e$=$e$=1), although
energies are expressed in Rydbergs (Ry). The magnetic field $B$ is
given in a.u. with $B_0= 2.35 \times 10^9\,G$ \footnote{In the
absence of convention, majority of results presented in literature
are obtained for $B_0= 2.35 \times 10^9\,G$, although sometimes
another convention is used $B_0= 2.3505 \times 10^9\,G$. Thus, in
making a comparison of the high accuracy results obtained by
different authors especially for high magnetic fields this fact
should be taken into account}.

\section{Generalities}

The Hamiltonian which describes two infinitely heavy protons and
one electron placed in a uniform constant magnetic field directed
along the $z-$axis, ${\bf B}=(0,0,B)$ is given by (see e.g.
\cite{LL})
\begin{equation}
\label{Ham}
 {\cal H} = {\hat p}^2 + \frac{2}{R} -\frac{2}{r_1} -\frac{2}{r_2}
  + ({\hat p} {\cal A}+{\cal A}{\hat p}) +  {\cal A}^2 \ ,
\end{equation}
(see Fig.1 for notations), where ${\hat p}=-i \nabla$ is the
momentum, ${\cal A}$ is a vector potential  which corresponds to
the magnetic field $\bf B$ and is chosen in the symmetric gauge to
be
\[
   {\cal A}= \frac{B}{2} (-y,\ x,\ 0)\ .
\]
Hence  the total energy $E_T $ of $H_2^+$ is defined as the total
electronic energy plus the Coulomb energy of proton repulsion. In
turn, the binding energy is defined as an affinity to having the
electron as well as both protons being infinitely separated,
$E_b=B-E_T$. The dissociation energy is defined as an affinity to
having one proton at infinity, $E_d=E_H-E_T$, where $E_H$ is the
total energy of the hydrogen atom in a magnetic field $B$. Spin
degrees of freedom can be separated out and their analysis is
straightforward.

The problem is characterized by two integrals of motion: (i)
angular momentum projection on the magnetic field direction
($z$-direction) and (ii) spatial parity $p$. Sometimes the parity
$\si$ corresponding to interchange of charged centers $1
\leftrightarrow 2$ is used, which is connected with the magnetic
quantum number and spatial parity,
\[
   p = \si (-1)^{|m|}\ .
\]
If the case $m$ is even, both parities coincide, $p=\si$. Thus,
any eigenstate has two definite quantum numbers: the magnetic
quantum number $m$ and the parity $p$ with respect $z \rar -z$.
Therefore the space of eigenstates is split into subspaces
(sectors) each of them is characterized by definite $m$ and $\si$.
Notation for the state we are going to use is based on the
following convention: the first number corresponds to the number
of excitation - "principal quantum number", e.g. the number 1 is
assigned to the ground state, then a Greek letter $\si, \pi, \de$
corresponds to $m=0,-1,-2$, respectively, with subscript $g/u$
(gerade/ungerade) describing positive/negative parity with respect
$z \rar -z$.

Most of the excited states we study are the lowest states (of the
type of the ground state) of the sectors with different magnetic
quantum numbers $m$ and $p$. It is quite obvious from the physical
point of view that the ground states of the sectors with $m>0$
always have larger total energies than those with $m \leq 0$.
Therefore we restrict our consideration to the states with
$m=0,-1,-2$.

Conservation of the $z$-component of the angular momentum assumes
the wave function of the electron (in cylindrical coordinates
$(\rho,\varphi,z)$) can be taken in the representation:
\[
\Psi = \,e^{i m \varphi}\rho^{|m|}\,\psi_m \ ,
\]
where $m$ is magnetic quantum number. Let us gauge rotate the
Hamiltonian (\ref{Ham}),
\begin{equation}
\label{Ham_m}
 {\cal H}_m\ =\
 e^{-i m \varphi}\rho^{-|m|}\, {\cal H} e^{i m \varphi}\rho^{|m|}\
 = {\hat p_m}^2 + \frac{2}{R} -\frac{2}{r_1} -\frac{2}{r_2}
  + m B +  \frac{B^2 \rho^2}{4} \ ,
\end{equation}
where
\[
 {\hat p_m} = e^{-i m \varphi}\rho^{-|m|}\ {\hat p}\ e^{i m
 \varphi}\rho^{|m|}\ ,
\]
is the gauge rotated momentum (covariant momentum). The constant
term $mB$ describes the linear Zeeman effect splitting. It can be
absorbed to a definition of total energy. The representation
(\ref{Ham_m}) is rather convenient since each Hamiltonian for
fixed $m$ describes the family of eigenstates with quantum number
$m$ and can be treated independently of the states with $m'$
different from $m$. Now the Hamiltonian (\ref{Ham_m}) has only the
invariance corresponding to the spatial parity conservation.

\begin{figure}[tb]
\begin{center}
     \includegraphics*[width=3in,angle=-90]{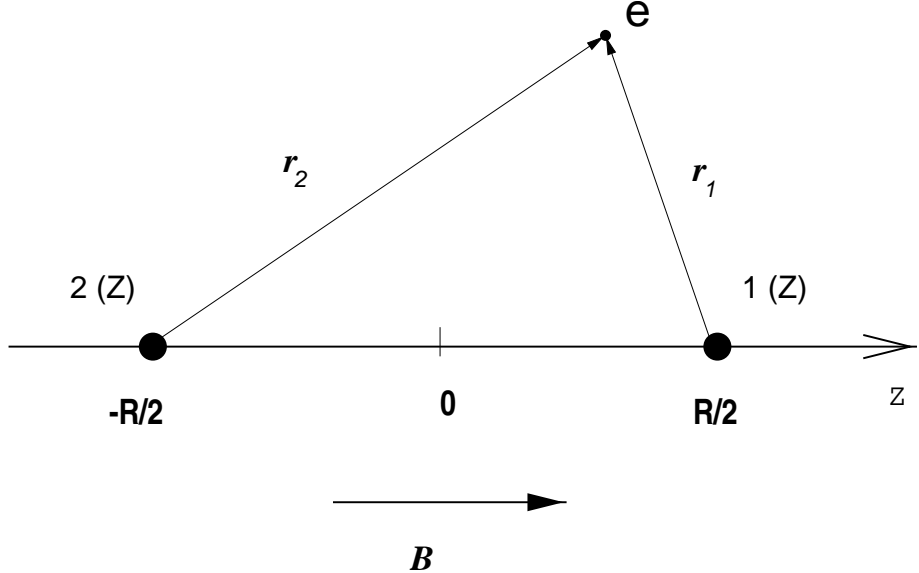}
     \caption{Geometrical setting  for the $H_2^{+}$ ion placed in a
       magnetic field directed along the $z$-axis. The protons are
       situated in the $y-z$ plane at a distance $R$ from each other
       and marked by bullets.}
    \label{fig:1}
\end{center}
\end{figure}

We are going to use the variational method in a way similar to
what was done in I. The recipe of choice of trial function is
based on physical arguments and is described in full generality in
\cite{Tur} (see for details the article I). Eventually, the ground
state trial function for fixed $m$ and $\si$ is chosen in a form
\begin{equation}
\label{anzatz}
  \psi_m^{(trial)} = A_1\psi_1 + A_2 \psi_2 + A_3\psi_3\, ,
\end{equation}
where
\begin{eqnarray}
\label{psi123}
  \psi_1 &=& \left\{\begin{array}{ll}
{e}^{-\al_1 (r_1+r_2)} {e}^{-B\be_{1}\rho^2 }\, , &
          \qquad \mbox{if}\ \si=+1\, ,\\
0 \, , &  \qquad \mbox{if}\ \si=-1\, , \\
\end{array}
\right .
\non \\
  \psi_2 &=& \big({e}^{-\al_2 r_1} + \si {e}^{-\al_2 r_2}\big)
{e}^{ -  B  \be_{2} \rho^2 }\, ,
\non \\
  \psi_{3} &=& \big({e}^{-\al_3 r_1-\al_4 r_2} + \si
{e}^{-\al_3 r_2-\al_4 r_1}\big) {e}^{ - B  \be_{3} \rho^2 }\ ,
\non
\end{eqnarray}
and $\si = \pm 1, \ m=0, \pm 1, \pm 2 \ldots$. Here $A_{1,2,3}$
and $\al_{1,2,3,4}, \beta_{1,2,3}$ as well as $R$ are variational
parameters, which are certainly different for different $m$
\footnote{Due to normalization of wave function one of $A$'s
should be kept fixed. Usually, we put $A_3=1$}. The functions
$\psi_{1,2,3}$ carry a certain physical meaning. They describe
coherent (incoherent) interaction of the electron with the protons
as well as their non-linear interpolation, respectively.
Calculations were performed  using the minimization package MINUIT
from CERN-LIB. Numerical integrations were carried out with a
relative accuracy of $\sim 10^{-9}$ by use of the adaptive NAG-LIB
(D01FCF) routine. All calculations were performed on dual PC
Pentium-4, with two processors of $2.8$ GHz each. Every particular
calculation of given eigenstate at fixed magnetic field including
minimization has taken in total about an hour of CPU time.
However, when the variational parameters are found it take a few
seconds of CPU time to calculate the variational energy.

It is necessary to mention two technical difficulties we
encountered. Calculation of two-dimensional integrals with high
accuracy which appeared in the problem has required a development
of a very sophisticated numerical technique. We created a
`dynamical partitioning' of the domain of integration, which
depend on values of variational parameters similar to what was
done in I. The domain partitioning was changed with a change of
the parameters. Sometimes the number of sub-domains was around 50.
Another technical problem is related with very complicated profile
of variational energy as the function of variational parameters
which is characterized by many local minima, saddle points and
valleys. Localization of the global minimum numerically of such a
complicated function with high accuracy is difficult technical
problem which becomes even more difficult in the case of ten or
more variational parameters. Examining the physical relevance of
trial functions allows one to avoid spurious minima. The
parameters obtained in (\ref{anzatz}) at every step of
minimization were always examined from the physical point of view.
Such considerations are always something of an art.

\section{$m=0$ }

The $m=0$ subspace consists of two subspaces, $\si=1$ (even
states) and $\si=-1$ (odd states).

\subsection{$1\si_g$ state ($\si = 1$)}

The state $1_g$ was thoroughly investigated in the paper I for the
whole range of inclinations $\tha=0^o - 90^o$ (for settings see
below Fig.2). At $\tha=0^o$ this state becomes the state $1\si_g$
and our variational anzatz $\psi_{1\si_g}$ (\ref{anzatz})
describing this state depends on ten parameters. As was mentioned
above, the search for the global minimum numerically with high
accuracy in the case of so many variational parameters is a
difficult technical task. Although this state was thoroughly
studied in \cite{Lopez:1997} we decided to repeat the calculations
using a more sophisticated strategy for localizing the minimum.
The essential new element of the strategy was to impose an extra
(natural) condition that the variational parameters change
smoothly with $B$. Finally, it led to an improvement of the
results in comparison to I and to previous calculations. It is
worth mentioning that this recalculation is very important for
calculation of the excited $2\si_g$ state, where the orthogonality
condition on trial functions must be imposed,
$(\psi_{1\si_g},\psi_{2\si_g})=0$. It is evident that an intrinsic
inaccuracy in $\psi_{1\si_g}$ due to the approximate nature of the
trial function (\ref{anzatz}) as a function of $x$ is a source of
inaccuracy in the energy of the $\psi_{2\si_g}$ state. Thus, a
reduction of this inaccuracy requires knowledge of the function
$\psi_{1\si_g}$ as accurately as possible. The above-mentioned
strategy allowed us to improve our previous results reported in
\cite{Lopez:1997} on total and binding energies (see Table I) and
also on lowest rotational-vibrational energies (see Table II).
Qualitative conclusions obtained in \cite{Lopez:1997} remain
unchanged.

\begin{table*}
\caption{\label{Table:1}
  Total $E_T$ and binding $E_b$ energies, and equilibrium distance $R_{eq}$
  for the state $1\si_g$, which is the global ground state of the $H_2^+$
  ion. Error bars for the equilibrium distance
  indicate a domain in $R$ where the value of binding energy remains the same
  within the indicated number of their digits shown in the table.}
\begin{ruledtabular}
\begin{tabular}{lcccc}
    $B$ &  $E_T$ (Ry) &  $E_{b}$ (Ry) & $R_{eq}$ (a.u.)   \\
    \hline
  $ B=0 $
          & -1.20525  & 1.20525    & 1.9971      & Present       \\
  $ 10^{9}$\,G
          & -1.15070  & 1.57623    & 1.923 $\pm$ 0.003  & Present \\
         & -1.15072  & 1.57625    & 1.924  &  Wille \cite{Wille:1988}\\
  $ 1$\,a.u.
          & -0.94992  & 1.94992    & 1.752 $\pm$ 0.003 & Present \\
          & ---       & 1.9498     & 1.752  &  Larsen \cite{Larsen}\\
          & -0.94642  & 1.94642    & 1.76   &  Kappes et al \cite{Schmelcher}\\
  $ 10^{10}$\,G
          & 1.09044   &  3.16488   & 1.246 $\pm$ 0.002 & Present \\
          & 1.09031   &  3.16502   & 1.246  &  Wille \cite{Wille:1988}\\
  $ 10$\,a.u.
          & 5.65024   &  4.34976   & 0.957 $\pm$ 0.002  & Present \\
          & ---       &  4.35      & 0.950  & Wille \cite{Wille:1988}\\
          & ---       &  4.35      & 0.958  & Larsen \cite{Larsen}\\
          & ---       &  4.3346    & 0.950  & Vincke et al \cite{Vincke}\\
  $ 10^{11}$\,G
          & 35.0432   & 7.5100     & 0.593 $\pm$ 0.001  & Present \\
          & 35.0428   & 7.5104     & 0.593  &  Wille \cite{Wille:1988}\\
  $ 100$\,a.u.
          & 89.7090   & 10.2910    & 0.448 $\pm$ 0.001  & Present \\
          & ---       & 10.2892    & 0.446  & Wille \cite{Wille:1988}\\
          & ---       & 10.270     & 0.448  & Larsen \cite{Larsen}\\
          & ---       & 10.2778    & 0.446  & Vincke et al \cite{Vincke}\\
  $ 10^{12}$\,G
          & 408.3894  & 17.1425    & 0.283 $\pm$ 0.001  & Present \\
          & ---       & 17.0588    & 0.28   & Lai et al \cite{Salpeter:1992}\\
  $ 1000$\,a.u
          & 977.2214  & 22.7786    & 0.2197 $\pm$ 0.0005 & Present \\
          & ---       & 22.7694    & 0.219  & Vincke et al \cite{Vincke}\\

  $ 10^{13}$\,G
          & 4219.565  & 35.7539    & 0.1472 $\pm$ 0.0002 & Present \\
          & ---       & 35.74      & 0.15   & Lai et al \cite{Salpeter:1992}\\
  $ 10000$\,a.u
          & 9954.203  & 45.7972    & 0.1183 $\pm$ 0.0002 & Present \\
  $ 4.414\times{10}^{13}$\,G
          & 18728.477  & 54.5018   & 0.1016 $\pm$ 0.0002 & Present \\
\end{tabular}
\end{ruledtabular}
\end{table*}


\begin{table*}
  \caption{\label{Table:2}
    Energies of the lowest vibrational $(E_{vib})$ and rotational
    $(E_{rot})$ electronic states  associated with the $1\si_g$ state. }
\begin{ruledtabular}
\begin{tabular}{lccl}
\( B \) & \( E_{vib} \) (Ry)& \( E_{rot} \) (Ry)&
\\ \hline
 \( 10^{9} \) G
               & 0.011 & 0.0053 & Present \\
                & 0.011 & 0.0038 & Wille \cite{Wille:1987} \\
\( 1 \) a.u.
               & 0.014 & 0.0110 & Present \\
                & 0.014 & 0.0091 & Larsen    \cite{Larsen} \\
               & 0.014 & 0.0238 & Le Guillou et al (b)
               \cite{Legui}\\
\( 10^{10} \) G
               & 0.026 & 0.0408 & Present \\
               & 0.026 & 0.0308 & Wille \cite{Wille:1987}\\
\( 10 \) a.u.
                & 0.040 & 0.0790 & Present \\
               & 0.040 & 0.133  & Larsen\cite{Larsen} \\
                & 0.040 & 0.0844 & Le Guillou et al (b)
              \cite{Legui}\\
\( 10^{11} \) G
               & 0.085 & 0.2151 & Present \\
\( 100 \) a.u.
               & 0.132 & 0.4128 & Present \\
               & 0.141 & 0.365  & Larsen\cite{Larsen}\\
               & 0.13  &  ---   & Wunner et al
               \cite{Wunner:82}\\
              & 0.132 & 0.410  & Le Guillou et al (b)
               \cite{Legui}\\
\( 10^{12} \) G
              & 0.266 & 1.0926 & Present \\
              & 0.198 & 1.0375 & Khersonskij \cite{Kher3} \\
\( 1000 \) a.u.
            & 0.390 & 1.9273 & Present \\
            & 0.38  & 1.77   &  Larsen \cite{Larsen} \\
            & 0.39  &  ---   & Wunner et al \cite{Wunner:82}\\
            & 0.388 & 1.916  & Le Guillou et al (b) \cite{Legui}\\
\( 10^{13} \) G
               & 0.714 & 4.875  & Present \\
              & 0.592 & 6.890  & Khersonskij \cite{Kher3}  \\
\( 10000 \) a.u.
               & 0.993 & ---   & Present \\
\( 4.414\times 10^{13} \)G
              & 1.248 & 12.065 & Present\\
\end{tabular}
\end{ruledtabular}
\end{table*}
\clearpage

\subsection{$2\si_g$ state ($\si=1$)}

This is the first excited state in the family of states with
quantum numbers $m=0, \si=1$. Here the trial function is taken in
the form
\begin{equation}
\label{anzatz-excited}
 \psi_{\psi_{2\si_g}} = \tilde A_1 \psi_1 + \tilde A_2 \psi_2 + \tilde A_3 \psi_3 \, ,
\end{equation}
with
\[
\psi_1 = (r_1 + r_2 - {\cal C}_1) e^{-\tal_1 (r_1+r_2)-\tbe_1 B
  \rho^2/4}\, ,
\]
\[
\psi_2=\Big[(r_1-{\cal C}_2)e^{-\tal_2 r_1} + (r_2-{\cal
C}_2)e^{-\tal_2 r_2}\Big]e^{-\tbe_2 B \rho^2/4} \, ,
\]
\[
\psi_3= \Big[(r_1 + a r_2 -{\cal C}_3)e^{-\tal_3 r_1 -\tal_4 r_2}
+ (r_2 + a r_1 -{\cal C}_3)e^{-\tal_3 r_2 -\tal_4 r_1}
\Big]e^{-\tbe_3 B \rho^2/4} \, ,
\]
(cf. (\ref{anzatz})), where $\tilde A_{1,2,3}$ and
$\tal_{1,2,3,4}, \tbe_{1,2,3}, a, {\cal C}_{1,2,3}$ as well as $R$
are variational parameters \footnote{Due to normalization of wave
function one of $\tilde A$'s can be kept fixed. Usually, we put
$\tilde A_3=1$}. This eigenfunction should be orthogonal to the
$\psi_{1\si_g}$ state trial function found in the previous
Section. The total number of variational parameters in
(\ref{anzatz-excited}) is 13.

The results obtained are presented in Table III. This state is
characterized by much smaller binding energy compared to the
$1\si_g$ state and is much more extended. The binding energy
displays a rather slow increase while the equilibrium distance
decreases slowly as the magnetic field grows. This excited state
is unstable with respect to dissociation to $H+p$.

\begin{table*}
\label{T2sg}
  \caption{Total $E_T$ and binding $E_b$ energies, and equilibrium distance $R_{eq}$
  for the state $2\sigma_g$ ($m=0,\sigma=1$).  }
\begin{ruledtabular}
\begin{tabular}{lllll}
    $B$ &  $E_T$ (Ry) &  $E_{b}$ (Ry) & $R_{eq}$ (a.u.) &  \\
    \hline
  $ B=0 $
          & -0.350032  & 0.350032  & 8.8   &  Present \\
          & -0.34936   & ---       & 8.8   &  Kappes \cite{Schmelcher}\\
          & -0.350098  & ---       & 8.834 &  Peek-Katriel \cite{Peek:1980}\\
  $ 10^{9}$\,G
          & -0.121343  &  0.546875  & 7.55   &  Present\\
          & -0.081824  & ---        & 7.792  &  Peek-Katriel \cite{Peek:1980}\\
  $ 1$\,a.u.
          &  0.34912   & 0.65088  & 6.640  &  Present\\
          &  0.34918   & 0.65082  & 6.632  &  Alarcon et al \cite{Alarcon:2003}\\
          &  0.34928   &  ---     & 6.64   &  Kappes et al \cite{Schmelcher}\\
  $ 10^{10}$\,G
          & 3.39938    & 0.85594  & 5.2  &  Present\\
  $ 10$\,a.u.
          & 9.02452    & 0.97548  & 4.6  & Present\\
  $ 10^{11}$\,G
          & 41.4090    & 1.1442   & 3.91 &  Present\\
  $ 100$\,a.u.
          & 98.7822    & 1.2178   & 3.65 & Present\\
  $ 10^{12}$\,G
          & 424.2277   & 1.3042   & 3.40 & Present\\
  $ 1000$\,a.u
          & 998.6620  & 1.3380    & 3.30 & Present\\
  $ 10^{13}$\,G
          & 4253.937  & 1.382     & 3.21 & Present\\
  $ 10000$\,a.u
          & 9998.608  & 1.392     & 3.145 & Present\\
  $ 4.414\times{10}^{13}$\,G
          & 18781.576 & 1.402     & 3.120  & Present\\
\end{tabular}
\end{ruledtabular}
\end{table*}

\clearpage

\subsection{$1\si_u$ state ($\si=-1$)}

In the absence of a magnetic field the $1\si_u$ state ($m=0,
\sigma=-1)$ is essentially repulsive and antibonding. However, in
a strong magnetic field this state becomes bound. Due to this fact
we want to study this state in full generality, for different
magnetic fields and inclinations.

In the absence of a magnetic field, the $1\si_u$ state is
characterized by a shallow minimum in the total energy situated at
large internuclear distance (see, for example,
\cite{Schmelcher}(a), \cite{Peek:1980}). Also this state is a
weakly bound state with respect to dissociation and it becomes
even unbound if nuclear motion is taken into account. So far  not
many studies have been carried out for this state. Our major
finding is that in the presence of a magnetic field of the
magnitude $ 10^9 < B \lesssim 4.414 \times 10^{13}$\,G the total
energy surface of the system $(ppe)$ in the state $1\si_u$
exhibits a well-pronounced minimum. Similar to the $1\si_g$ state,
both total ($E_T$) and binding ($E_b$) energies of the $1\si_u$
state increase as the magnetic field grows, while the equilibrium
distance decreases. However, the accuracy of our calculations does
not allow us to make a definitive conclusion about the stability
of the system in this state with respect to dissociation and
nuclear motion effects. In the case of non-zero inclination $\tha
\neq 0^o$ (for definition see Fig. 2) we denote this state as
$1_u$ reflecting the fact that the only parity conservation
exists. In I it was shown that for $B \gtrsim 10^{11}\,G$ and
large inclinations the $1_g$ state disappears and hence the
molecular ion $H_2^{+}$ does not exist. Thus, it seems it makes no
sense to study the $1_u$ state in this domain. We checked a
consistency of this statement verifying that always inequality
$E_T^{1\si_g} (R) < E_T^{1\si_u} (R)$ holds.

To study the $1_u$ state we use the following form of the vector
potential corresponding to a constant magnetic field ${\bf
B}=(0,0,B)$
\begin{equation}
\label{Vec}
  {\cal A}= B((\xi-1)y,\ \xi x,\ 0)\ ,
\end{equation}
where $\xi$ is a parameter, which later will be considered as
variational. If $\xi=1/2$ we get the well-known and widely used
gauge which is called symmetric or circular. If $\xi=0$ or 1, we
get the asymmetric or Landau gauge (see \cite{LL}). By
substituting (\ref{Vec}) into (\ref{Ham}) we arrive at a
Hamiltonian of the form
\begin{widetext}
\begin{eqnarray}
  \label{Ham.fin}
 {\cal H} = -{\nabla}^2 + \frac{2}{ R} -\frac{2}{r_1}
 -\frac{2}{r_2}
 -  2 i B[(\xi-1) y \pa_x  + \xi x \pa_y]
+  B^2 [ \xi^2 x^2+ (1-\xi)^2 y^2 ] \ .
\end{eqnarray}
\end{widetext}

\begin{figure}
\includegraphics*[width=2.0in,angle=-90]{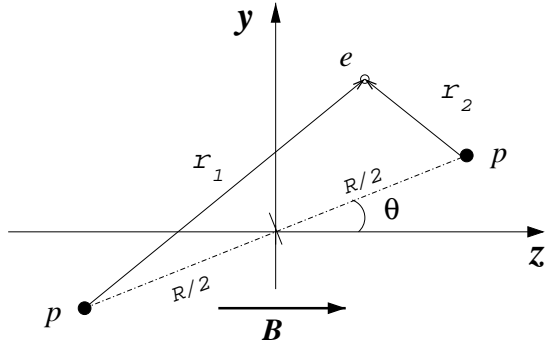}%
\caption{\label{fig:2} Geometrical setting  for the inclined
H$_2^+$ ion in a magnetic field directed along the $z$-axis.}
\end{figure}

The trial function is chosen in the form
\begin{equation}
\label{sigma_u}
 \psi_{1_u}= A_1 \psi_{1} + A_2 \psi_{2}\ ,
\end{equation}
with
\[
 \psi_1= \big({e}^{-\al_1 r_1} - {e}^{-\al_1 r_2}\big) {e}^{ -
B  [\be_{1x}\xi x^2 +\be_{1y}(1-\xi) y^2] }\ ,
\]
\[
 \psi_2= \big({e}^{-\al_2 r_1-\al_3 r_2} - {e}^{-\al_2
r_2-\al_3 r_1}\big) {e}^{ - B  [\be_{2x}\xi x^2 +\be_{2y}(1-\xi)
y^2] }\ ,
\]
where $A_1, A_2$ are parameters and one of them is kept fixed by a
normalization condition. All parameters $\al_{1,2,3},
\be_{1x,1y,2x,2y}, A_1, A_2$ and $\xi$ are variational parameters.
It is evident that if $\tha=0^o$, the rotational invariance along
$z$-axis exists and the vector potential should be taken in a form
supporting this invariance. Hence the parameter $\xi$ in
(\ref{Vec}) takes value $\xi=1/2$ and the parameters
$\be_{1x}=\be_{1y}, \be_{2x}=\be_{2y}$.

Numerical study for the $1_u$ state was carried out for different
inclinations with the results at $0^{\circ}, 45^{\circ}$ and
$90^{\circ}$ for magnetic fields $B=0 - 4.414 \times 10^{13}$ G as
shown in Tables IV-VI. The immediate conclusion is that
\[
E_T(0^{\circ})<E_T(45^{\circ})< E_T(90^{\circ})
\]
for all magnetic fields, where this comparison makes sense (see
below). Hence, similar to the $1_g$ state, the highest molecular
stability of the $1_u$ state occurs for the parallel
configuration, at $\tha=0^o$ (see I). Also, the binding energy
growth is maximal as a function of magnetic field for the parallel
configuration. Therefore, the stability of $H_2^+$ in the parallel
configuration in the $1_u$ state increases as the magnetic field
grows, again similarly to what happens for the $1_g$ state. These
results suggest the following picture for appearance of a bound
state for the $1_u$ state: for small magnetic fields the minimum
in the total energy arises at very large internuclear distances
 \footnote{It is natural to
  assume that for $B=0$ a minimum exists at infinite (or almost
  infinite) internuclear distance.},
then, as the  magnetic field grows, the position of the minimum
moves to smaller and smaller internuclear distances.

\begin{table*}
\label{T1su}
  \caption{$1\sigma_u$ state in the parallel configuration, $\tha=0^{\circ}$.
    Total ($E_T$) and binding ($E_b$) energies are in Ry and
    equilibrium distance \( R_{eq} \) in a.u.
}
\begin{ruledtabular}
    \begin{tabular}{lllll}
\( B \) &  \( E_T \) (Ry)& \( E_{b} \) (Ry)& \( R_{eq} \) (a.u.)& \\
\hline
$B=0$         & -1.00010   & 1.00010   & 12.746 & Lopez et al \cite{Lopez:1997} \\
              & -1.00012   & 1.00012   & 12.55   &Peek-Katriel
              \cite{Peek:1980}\\[5pt]
 \( 10^{9} \) G
              & -0.92103   & 1.34656   & 11.19  & Present \\
              & -0.917134  &  ---      & 10.55  & Peek-Katriel
              \cite{Peek:1980}\\[5pt]
 \( 1 \) a.u.
              &  -0.66271  & 1.66271   & 9.73   & Present\\
              &  -0.66     & 1.66      & 9.6    & Kappes et al
              \cite{Schmelcher}\\[5pt]
\( 10^{10} \) G
               &  1.63989  & 2.61500   & 7.18   & Present \\
               &  2.1294   &  ---      & 4.18   & Peek-Katriel
               \cite{Peek:1980}\\[5pt]
 \( 10 \) a.u.
               &  6.52362 & 3.47638  & 6.336   & Present\\[5pt]
 \( 10^{11} \)G
               & 36.8367  & 5.7165   & 4.629  & Present \\[5pt]
 \( 100 \) a.u.
               & 92.4257  & 7.5743   & 3.976  & Present\\[5pt]
 \( 10^{12} \)G
               & 413.6175 & 11.9144  & 3.209  & Present \\[5pt]
 \( 1000 \) a.u.
               & 984.6852 & 15.3148  & 2.862  & Present\\[5pt]
 \( 10^{13} \)G
               & 4232.554 & 22.765   & 2.360  & Present \\[5pt]
 \( 10000 \) a.u.
               & 9971.727 & 28.273   & 2.134  & Present\\[5pt]
\( 4.414\times 10^{13} \) G
               & 18750.07 & 32.912   & 2.021  & Present \\[5pt]
\end{tabular}
\end{ruledtabular}
\end{table*}

\begin{table}
  \caption{\label{Table:5}
  Total $E_T$, binding $E_b$ energies and equilibrium distance $R_{eq}$
  for the $1_u$ state in the configuration $\tha=45^{\circ}$.
  Optimal value for the gauge  parameter $\xi$ is shown (see text).}
\begin{ruledtabular}
    \begin{tabular}{lcccc}
\( B \)&  \( E_T \) (Ry)& \( E_{b} \) (Ry)& \( R_{eq} \) (a.u.)&\( \xi  \)\\
\hline
 \(10^{9}\)G
             & -0.870391 & 1.295923  & 8.053   & 0.9308  \\[5pt]
 \( 1 \) a.u.
             & -0.509041 & 1.509041  & 6.587   & 0.9406  \\[5pt]
 \(10^{10}\) G
             &  2.267998 & 1.987321  & 4.812   & 0.9671  \\[5pt]
 \( 10 \) a.u.
             &  7.692812 & 2.307188  & 4.196   & 0.9808  \\[5pt]
  \(10^{11}\) G
             & 39.71061  & 2.84258   & 3.538   & 0.9935  \\[5pt]
  \( 100 \) a.u.
             & 96.88464  & 3.11536   & 3.278   & 0.9968  \\[5pt]
  \(10^{12}\) G
             & 422.0074  & 3.5245    & 3.020   & 0.9991  \\[5pt]
  \( 1000 \) a.u.
             & 996.3044  & 3.6956    & 2.894   & 0.9996  \\[5pt]
\end{tabular}
\end{ruledtabular}
\end{table}

\begin{table*}
  \caption{\label{Table:6}
  Total $E_T$, binding $E_b$ energies and equilibrium distance
  $R_{eq}$  for the $1_u$ state at $\tha=90^{\circ}$.  Optimal value
  for the gauge   parameter $\xi$ is shown (see text). }
\begin{ruledtabular}
    \begin{tabular}{lccccl}
\( B \)&  \( E_T \) (Ry)& \( E_{b} \) (Ry)& \( R_{eq} \) (a.u.)&\( \xi  \)& \\
\hline
\(10^{9}\) G
 & -0.867234  & 1.292766  & 8.784  & 0.9692  & Present  \\[5pt]
\(1\) a.u.
 & -0.49963   & 1.49963   & 7.264  & 0.9737  & Present  \\
 & -0.65998   & 1.65998   & 5.45   &         & Kappes et al
 \cite{Schmelcher}(b)  \\[5pt]
\(10^{10}\) G
 &  2.29365   & 1.96167   & 5.517  & 0.9866  & Present  \\[5pt]
\(10\) a.u.
 &  7.72998   & 2.27002   & 4.872  & 0.9923  & Present  \\[5pt]
\(10^{11}\) G
 & 39.76500   & 2.78819   & 4.154  & 0.9975  & Present  \\[5pt]
\(100\) a.u.
 & 96.93497   & 3.06503   & 3.875  & 0.9988  & Present  \\[5pt]
\end{tabular}
\end{ruledtabular}
\end{table*}


Our results for $B>0$ and $\tha=0^{\circ}$ give the lowest total
energies compared to other calculations. In general, they are in a
good agreement with those by Kappes--Schmelcher
\cite{Schmelcher}$^{(a)}$ as well as by Peek--Katriel
\cite{Peek:1980} for $B=0, 10^9$ G, although for $B=10^{10}$ G a
certain disagreement is observed (see Table IV). However, for
$\tha=90^{\circ}$ our results are in striking, qualitative
contrast with those by Wille \cite{Wille:1988}, where even the
optimal configuration is attached to $\tha=90^{\circ}$, contrary
to our conclusion. For instance, at $B=10^{10}$ G in
\cite{Wille:1988} the values $E_b=2.593$ Ry and $R_{eq}=2.284$
a.u. are given, while our results are $E_b=1.9617$ Ry and
$R_{eq}=5.517$ a.u., respectively (see Table VI). Similar, but
less drastic disagreement, is observed with the results in
\cite{Schmelcher}$^{(b)}$.  We can only guess this disagreement is
due to the shallow nature of the minimum, but a real explanation
of this fact is missing. Independent calculations are needed in
order to resolve this contradiction.

The analysis of Tables IV-VI shows that for $\tha>0^{\circ}$ and
fixed magnetic field the total energy of $H_2^+$ in the $1_u$
state is always larger than the total energy of the hydrogen atom
\cite{Kravchenko}.  It means that the $H_2^+$-ion in the $1_u$
state is unstable towards dissociation to $H + p$.  For $\tha\sim
0^{\circ}$ the total energies presented for the $H_2^+$ ion and
the most accurate results for the hydrogen atom \cite{Kravchenko}
are comparable within the order of magnitude $10^{-4} - 10^{-5}$.
We estimate that the accuracy of our calculations is of the same
order of magnitude $10^{-4} - 10^{-5}$. This prevents us from
making a conclusion about the stability of $H_2^+$ in the $1_u$
state with respect to dissociation. Thus, the only reliable
conclusion can be drawn that the minimum is very shallow.

The $1_u$ state is much more extended than the $1_g$ state: for fixed
magnetic field the equilibrium distance of the $1_g$ state is much
smaller than that for the $1_u$ state. This picture remains the same
for any inclination. It is quite impressive to observe the much lower
rate of decrease of $R_{eq}$ in the range $B=0 - 4.414 \times 10^{13}$
G with magnetic field growth. For example, in the case of the parallel
configuration, $\tha=0^o$, for the state $1_u$ the equilibrium
distance falls $\sim 6$ times compared to the $1_g$ state, where it
falls $\sim 20$ times.

The behavior of the equilibrium distance $R_{eq}$ of the $1_u$
state as a function of inclination is quite non-trivial (see
Tables IV-VI). As in the $1_g$ state, the H$_2^+$-ion in the $1_u$
state for $B \lesssim 10^{12}$ G is most extended in the parallel
configuration.

\section{$m=-1$}

The subspace consists of two subspaces, $\si=1$ (even states) and
$\si=-1$ (odd states).

\subsection{$1\pi_u$ state ($\si = 1$)}

In order to study the $1\pi_u$ ($m=-1$ and $\si=1$) state we take
the trial function (\ref{anzatz}). The results are presented in
Table VII. In general, our results are more accurate than those
obtained in other calculations giving lower total (and
correspondingly, the higher binding) energies with the only
exception of the magnetic field $B=10$\,a.u. where the result for
binding energy from \cite{Wille:1988} is better in the fourth
digit. The results for $B=1$ a.u. obtained in \cite{Wille:1988}
are not very precise in $R_{eq}$ (see Table I therein), which
explain their difference with the results by others. The binding
energy at $B=10$ and $100$ a.u. in \cite{Wille:1988} is calculated
for the same equilibrium distances as those found in \cite{Vincke}
(see Table IV in  Ref. \cite{Wille:1988}). Like for all studied
states the binding energy grows steadily with magnetic field
increase while the equilibrium distance shrinks in a quite drastic
manner. If for small magnetic fields the equilibrium distance
$R_{eq}$ is several times larger than this distance for the
$1\si_g$ state, for large magnetic fields these equilibrium
distances become comparable. Among $m=-1$ states the state
$1\pi_u$  has the smallest total energy.

\begin{table*}
\label{T1pu}
    \caption{Total $E_T$, binding $E_b$ energies and equilibrium distance $R_{eq}$
  for the excited state $1\pi_u$ ($m=-1, \si = 1$).}
\begin{ruledtabular}
\begin{tabular}{lllll}
$B$            & $E_T$ (Ry)  &  $E_b$ (Ry) & $ R_{eq}$(a.u.)&  \\
\hline
$10^9$\, G     & -0.293592   & 0.719123    & 4.940   & Present\\
$1$\, a.u.     & -0.020150   & 1.020150    & 3.676   & Present\\
               & -0.02014    &    --       & 3.68    & Kappes et al
               \cite{Schmelcher} \\
               & -0.02011    &   --        & 3.75    & Wille \cite{Wille:1988} \\
$10^{10}$\, G  & 2.371845    & 1.883474    & 2.130   & Present\\
$10$\, a.u.    & 7.29682     & 2.70318     & 1.526   & Present\\
               &             & 2.6862      & 1.510   & Vincke-Baye \cite{Vincke} \\
               &             & 2.7046      & 1.510   & Wille \cite{Wille:1988} \\
$10^{11}$\, G  & 37.6490     & 4.9042      & 0.887   & Present\\
$100$\, a.u.   & 93.1127     & 6.8873      & 0.651   & Present\\
               &             & 6.8774      & 0.645   & Vincke-Baye \cite{Vincke} \\
               &             & 6.8548      & 0.645   & Wille \cite{Wille:1988} \\
$10^{12}$\, G  &  413.6306   & 11.902      & 0.395   & Present\\
$1000$\, a.u.  &  983.874    & 16.126      & 0.301   & Present\\
$10^{13}$\, G  & 4229.183    & 26.136      & 0.195   & Present\\
$10000$\, a.u. & 9965.932    & 34.068      & 0.154   & Present\\
$4.414\times 10^{13}$\, G
               & 18741.89    & 41.09       & 0.130    & Present\\
    \end{tabular}
\end{ruledtabular}
\end{table*}
\clearpage

\subsection{$1\pi_g $ state ($\si=-1$)}

In order to study the $1\pi_g$ state ($m=-1$ and $\si=-1$) we take
the trial function (\ref{anzatz}). The results are presented in
Table VIII. For $B=1$\,a.u. our total energy deviates from
\cite{Schmelcher} in the third digit and an independent
calculation would be desirable.

\begin{table*}
\label{T1pg}
    \caption{Total $E_T$, binding $E_b$ energies and equilibrium distance $R_{eq}$
  for the state $1\pi_g$ ($m=-1, \si =-1$).}
\begin{ruledtabular}
\begin{tabular}{lllll}
$B$            & $E_T$ (Ry)  &  $E_b$ (Ry)&$R_{eq}$(a.u.)  &   \\
\hline
$10^9$\, G     &-0.232060    &  0.65759   & 20.10   & Present \\
$1$\, a.u.     & 0.086868    &  0.91313   & 14.05   & Present \\
               & 0.0866      &            & 13.5      & Kappes et al
               \cite{Schmelcher} \\
$10^{10}$\, G  & 2.641122    & 1.61420    & 9.370   & Present \\
$10$\, a.u.    & 7.749819    & 2.25018    & 7.622    & Present \\
$10^{11}$\, G  & 38.67642    & 3.87677    & 5.622    & Present \\
$100$\, a.u.   & 94.73386    & 5.26614    & 4.791    & Present \\
$10^{12}$\, G  & 416.9354    & 8.59654    & 3.767  & Present \\
$1000$\, a.u.  & 988.7286    & 11.2714    & 3.321    & Present \\
$10^{13}$\, G  & 4238.038    & 17.2810    & 2.708   & Present \\
$10000$\, a.u. & 9978.175    & 21.8254    & 2.420   & Present \\
$4.414\times 10^{13}$\, G
                 & 18757.273  & 25.7054  & 2.237    & Present \\
\end{tabular}
\end{ruledtabular}
\end{table*}
\clearpage

\section{$m=-2$}

The subspace consists of two subspaces, $\si=1$ (even states) and
$\si=-1$ (odd states).

\subsection{$1\de_g$ state ($\si=1$)}

In order to study the $1\de_g$ state ($m=-2$ and $\si=1$) we take
the trial function (\ref{anzatz}). The results are presented in
Table IX. In \cite{Wille:1988} for $B=1$ a.u. the equilibrium
distance is simply placed equal to $5.0$\,a.u. (see Table I
therein). For $B=10,100$ a.u. the energies computed in
\cite{Wille:1988} were calculated for the same equilibrium
distances as those found in \cite{Vincke} (see Table IV in
\cite{Wille:1988}). Among $m=-2$ states the $1\de_g$ state has the
smallest total energy. It is worth mentioning a drastic decrease
of $R_{eq}$ with magnetic field growth similar to what appears for
$1\si_g$ and $1\pi_u$ states.

\begin{table*}
\label{T1dg}
    \caption{Total $E_T$, binding $E_b$ energies and equilibrium distance $R_{eq}$
   for the state $1\delta_g$ ($m=-2,\sigma=+1$).}
\begin{ruledtabular}
    \begin{tabular}{lllll}
$B$            & $E_T$ (Ry)  &  $E_b$ (Ry) &$R_{eq}$(a.u.) &   \\
\hline
$10^9$\, G     & -0.107945 & 0.533477  & 6.865   &  Present\\
$1$\, a.u.     & 0.221163  & 0.778837  & 4.872   &  Present\\
               & 0.22112   &           & 4.87    & Kappes et al
               \cite{Schmelcher} \\
               & 0.22126   &           & 5.0     & Wille \cite{Wille:1988}  \\
$10^{10}$\, G  & 2.77538   & 1.47994   & 2.694   &  Present\\
$10$\, a.u.    & 7.85113   & 2.14887   & 1.907   &  Present\\
               &           & 2.1306    & 1.880   & Vincke-Baye \cite{Vincke} \\
               &           & 2.1496    & 1.880   & Wille \cite{Wille:1988} \\
$10^{11}$\, G  & 38.58470  & 3.9685   & 1.080   & Present\\
$100$\, a.u.   & 94.38093  & 5.6191    & 0.782   & Present \\
               &           & 5.6058    & 0.778   & Vincke-Baye \cite{Vincke} \\
               &           & 5.510     & 0.778   & Wille \cite{Wille:1988} \\
$10^{12}$\, G  & 415.6710  & 9.8609    & 0.470   & Present \\
$1000$\, a.u.  & 986.5119  & 13.4881   & 0.353   & Present \\
$10^{13}$\, G  & 4233.125  & 22.194    & 0.225   & Present \\
$10000$\, a.u. & 9970.802  & 29.198    & 0.176   & Present \\
$4.414\times 10^{13}$\, G
               & 18747.572 & 35.407    & 0.148   & Present \\
    \end{tabular}
\end{ruledtabular}
\end{table*}

\clearpage

\subsection{$1\de_u$ state ($\si=-1$)}

In order to study the $1\de_u$ state ($m=-2$ and $\si=-1$) we take
the trial function (\ref{anzatz}). The results are presented in
Table X.

\begin{table*}
\label{T1du}
    \caption{Total $E_T$, binding $E_b$ energies and equilibrium distance $R_{eq}$
    for the state $1\de_u$ ($m=-2,\si=-1$).}
\begin{ruledtabular}
    \begin{tabular}{lllll}
$B$            & $E_T$ (Ry)  &  $E_b$ (Ry)&$R_{eq}$(a.u.)  &  \\
\hline
$10^9$\, G     & -0.06873  & 0.49426 & 23.902  &  Present \\
$1$\, a.u.     &  0.29410  & 0.70590 & 16.377  &  Present \\
               &  0.2936   & ---     & 16.0    &  Kappes et al
               \cite{Schmelcher} \\
$10^{10}$\, G  &  2.97742  & 1.27790 & 11.475  & Present \\
$10$\, a.u.    &  8.19892  & 1.80108 & 9.458   & Present \\
$10^{11}$\, G  & 39.40596  & 3.14723 & 6.858   & Present \\
$100$\, a.u.   & 95.69542  & 4.30458 & 5.619   & Present \\
$10^{12}$\, G  & 418.4335  & 7.0984  & 4.071   & Present\\
$1000$\, a.u.  & 990.6416  & 9.3584  & 3.406   & Present\\
$10^{13}$\, G  & 4240.834  & 14.485  & 2.625   & Present\\
$10000$\, a.u. & 9981.587  & 18.413  & 2.391   & Present\\
$4.414\times 10^{13}$\, G
               & 18761.18  & 21.80   & 2.230   & Present \\
    \end{tabular}
\end{ruledtabular}
\end{table*}
\clearpage

\section{Discussion}

In Table XI a summary of total energies of eigenstates explored in
this article for magnetic fields ranging from $10^9$\,G to $4.414
\times 10^{13}$\,G is presented. An analysis of Table IX allows to
draw a certain immediate conclusions:
\begin{enumerate}
    \item The state $1\si_g$ is the global ground state for all
    magnetic fields. It is rather evident that
    this statement remains valid in general, when even the states other
    than studied are taken into account (Perron theorem);
    \item For the states with fixed $m$ the lowest total energy
    corresponds to the state of positive parity $\si=+1$. We
    guess that this statement remains correct in general;
    \item For the same parity $\si$ ground states are ordered
    following the value of $m$,
\[
 E_T^{1\si_g} < E_T^{1\pi_u} < E_T^{1\de_g} \ ,
\]
\[
 E_T^{1\si_u} < E_T^{1\pi_g} < E_T^{1\de_u} \ .
\]
    \item There exist several true level crossings,
   \begin{itemize}
      \item For $B \lesssim 10^{12}$\,G
\[
 E_T^{1\si_u} < E_T^{1\pi_u}  \ ,
\]
while for $B > 10^{12}$\,G
\[
 E_T^{1\si_u} > E_T^{1\pi_u}  \ ,
\]
      \item For $B \lesssim 10 000$\,a.u.
\[
 E_T^{1\si_u} < E_T^{1\de_g}  \ ,
\]
while for $B > 10 000$\,a.u.
\[
 E_T^{1\si_u} > E_T^{1\de_g}  \ ,
\]
      \item For $B \lesssim 10$\,a.u.
\[
 E_T^{1\pi_g} < E_T^{1\de_g}  \ ,
\]
while for $B > 10$\,a.u.
\[
 E_T^{1\pi_g} > E_T^{1\de_g}  \ ,
\]
      \item For $B \lesssim 1$\,a.u.
\[
 E_T^{1\de_g} \geq E_T^{2\si_g}  \ ,
\]
while for $B > 1$\,a.u.
\[
 E_T^{1\de_g} < E_T^{2\si_g}  \ ,
\]
      \item For $B \lesssim 1$\,a.u.
\[
 E_T^{1\de_u} \geq E_T^{2\si_g}  \ ,
\]
while for $B > 1$\,a.u.
\[
 E_T^{1\de_u} < E_T^{2\si_g}  \ .
\]
   \end{itemize}
\end{enumerate}

\begingroup
\squeezetable
\begin{table*}[hb]
\label{TETs}
    \caption{Comparison of the total energies $E_T$ (in Rydbergs)
      for the low-lying states of the $H_2^+$ molecular ion for
      magnetic fields $10^9$\,G - $4.414\times 10^{13}$\,G.}
    \begin{ruledtabular}
    \begin{tabular}{lccccccc}
 $B$        &$1\sigma_g$ &$1\sigma_u$ & $1\pi_u$ & $1\pi_g$ & $1\delta_g$ & $1\delta_u$ &$2\sigma_g$  \\
\hline
$10^9$ \, G  & -1.15070  & -0.92103 & -0.29359 & -0.232060&-0.107945 &-0.068727 & -0.121343     \\[5pt]
$1$ \, a.u.  & -0.94992  & -0.66271 & -0.20150 & 0.086868 & 0.22117  & 0.29410  & 0.34912     \\[5pt]
$10^{10}$\,G & 1.09044   & 1.63989  &  2.371845& 2.641122 & 2.77538  & 2.977418 & 3.39938     \\[5pt]
$10$\,a.u.   & 5.65024   & 6.52362  & 7.296816 & 7.749819 & 7.85113  & 8.198922 & 9.02452     \\[5pt]
$10^{11}$\,G & 35.04320  & 36.83671 & 37.64895 & 38.67642 & 38.58470 & 39.40596 & 41.4090     \\[5pt]
$100$\, a.u. & 89.7090   & 92.4257  & 93.11267 & 94.7339  & 94.38093 & 95.69542 & 98.7822     \\[5pt]
$10^{12}$\,G & 408.3894  & 413.6175 & 413.6306 & 416.9354 & 415.6710 & 418.4335 & 424.2278     \\[5pt]
$1000$\, a.u.& 977.2214  & 984.685  & 983.874  & 988.7286 & 986.5119 & 990.6416 & 998.662     \\[5pt]
$10^{13}$\, G
           & 4219.565  & 4232.554 & 4229.183  & 4238.038 &4233.126 &4240.834 & 4253.937   \\[5pt]
$10000$\, a.u.& 9954.203  & 9971.727 & 9965.932 & 9978.175 & 9970.802 & 9981.587 & 9998.608  \\[5pt]
$4.414\times 10^{13}$\, G
     & 18728.477 & 18750.070 & 18741.889  & 18757.273 & 18747.572 & 18761.180 & 18781.576  \\[5pt]
    \end{tabular}
\end{ruledtabular}
\end{table*}
\endgroup
\clearpage

\section{Conclusion}

We have carried out an accurate, non-relativistic calculation in
the Born-Oppenheimer approximation for the low-lying states of the
$H_2^+$ molecular ion in the parallel configuration at equilibrium
in the framework of a unique computational approach. The $1\si_u$
state is considered in full generality for all inclinations of the
molecular axis vs. magnetic field direction. We studied constant
uniform magnetic fields ranging from $B=10^9\, G$ up to $B = 4.414
\times 10^{13}\,G$, where non-relativistic considerations hold,
although our method can be naturally applied to study the domain
$B<10^9 \,G$. We used a variational method with a very simple
trial function with a few variational parameters inspired by the
underlying physics of the problem. Thus our trial function can be
easily analyzed and in contrast to other approaches our results
can be easily reproduced. The trial function (3) can be easily
modified to explore other excited states.

The present study of several low-lying excited states complements
a study of the ground state performed in I. Usually the total,
binding, dissociation and transition energies grow with increase
in the magnetic field, reaching values of several hundred eV at
magnetic fields of $10^{12}-10^{13}$\,G. These results can be used
to construct a model of the atmosphere of an isolated neutron star
1E1207.4-5209 (see \cite{Sandal:2002}). This will be done
elsewhere.

\begin{acknowledgments}
  This work was supported in part by
  CONACyT grants {\bf 25427-E} and {\bf 36600-E} (Mexico).
\end{acknowledgments}

\end{document}